\documentclass[a4paper,12pt]{article}

\usepackage{amsmath,amssymb,amsfonts}
\allowdisplaybreaks
\usepackage{graphicx}
\usepackage{color}
\usepackage{caption}
\usepackage{subcaption}
\makeatletter
\@addtoreset{equation}{section}
\renewcommand{\theequation}{\thesection.\@arabic\c@equation}
\makeatother
\usepackage{hyperref}
\usepackage{cite}
\usepackage{caption}
\definecolor{red}{rgb}{1,0,0}
\definecolor{green}{rgb}{0,1,0}
\definecolor{blue}{rgb}{0,0,1}
\definecolor{darkblue}{rgb}{0,0,0.5}
\definecolor{lightblue}{rgb}{.5,.5,1}
\definecolor{lightgray}{gray}{.87}
\definecolor{Dark}{gray}{.20}
\definecolor{pink}{rgb}{.95,0.82,0.92}
\definecolor{yellow}{rgb}{1,1,0}
\definecolor{lightyellow}{rgb}{1,1,.5}
\definecolor{purple}{rgb}{0.7,0,0.85}
\definecolor{darkgreen}{rgb}{0,0.25,0}
\definecolor{darkpurple}{rgb}{0.4,0,0.4}
\definecolor{orange}{rgb}{0.8,0.2,0.2}
\def \be {\begin{equation}}
\def \ee {\end{equation}}
\def \bea {\begin{align}}
\def \eea {\end{align}}
\def \nn {\nonumber}
\def \la {\langle}
\def \ra {\rangle}
\def \rr {\raise.35ex\hbox{\small $\prime$}\kern-.17em{\mbox{\large $\imath$}}}
\def \del {\partial}
\def \dels {\partial\kern-.5em / \kern.5em}
\def \As {{A\kern-.5em / \kern.5em}}
\def \Ds {D\kern-.7em / \kern.5em}

\def \dag {\dagger}

\def \eps {\epsilon}

\def \s {\sigma}

\def \om {\omega}

\def \th {\theta}
\def \Th {\Theta}

\def \rh {\hat{r}}

\newcommand{\boxedeq}[1]{
  \vskip5pt
  \fcolorbox{blue}{white}{
      \begin{minipage}{5.7in}
      \color{black}
      \be #1\ee
      \vskip2pt
      \end{minipage}
    }
    \vskip10pt
}

\newcommand{\hide}[1]{}

\newcommand{\explanation}[1]{}

\newcommand{\UE}[1]
{}

\pdfoutput=1

\begin{document}

\begin{titlepage}
\vspace*{-10mm}   
\baselineskip 10pt   
\begin{flushright}   
\begin{tabular}{r} 
RIKEN-iTHEMS-Report-26
\end{tabular}   
\end{flushright}   
\baselineskip 24pt   
\vglue 10mm   

\begin{center}

\noindent
\textbf{\LARGE
Macroscopic Black-Hole Remnants\\
in a Nonlocal Field Theory:\\
Towards Hawking Radiation in SFT\\
}
\vskip10mm
\baselineskip 20pt

\renewcommand{\thefootnote}{\fnsymbol{footnote}}

{\large
Feng-Yin Cheng${}^{a}$\,\footnote[1]{luke091x@gmail.com},
Pei-Ming~Ho${}^{a,b}$\,\footnote[2]{pmho@phys.ntu.edu.tw},
Wei-Hsiang Shao${}^{c}$\,\footnote[3]{whsshao@gmail.com}
}
\vskip1em

\renewcommand{\thefootnote}{\arabic{footnote}}


{\it
${}^{a}$
Department of Physics and Center for Theoretical Physics, \\
National Taiwan University, Taipei 10617, Taiwan
\\
${}^{b}$
Physics Division, National Center for Theoretical Sciences, 
Taipei 10617, Taiwan
\\
${}^{c}$
RIKEN Center for Interdisciplinary Theoretical and Mathematical Sciences (iTHEMS), \\
RIKEN, Wako 351-0198, Japan
}

\vskip 20mm
\begin{abstract}
\normalsize

We demonstrate that, for a large black hole of radius $a$, Hawking radiation terminates around the scrambling time $u_{scr} \equiv 2a \log(a/\ell)$ due to the nonlocal, exponential suppression of trans-Planckian interactions inherent in string field theory (SFT). 
Modifying a massless scalar field's interaction with a dynamical black hole background via the smearing operator $e^{\ell^2\Box}$ (where $\ell$ denotes the string length scale), we calculate the time-dependent number expectation value $\langle \hat{N}(u) \rangle$ of outgoing Hawking particles at retarded time $u$. 
While the standard Planck spectrum at the Hawking temperature is reproduced at early times ($u \ll u_{scr}$), the particle number approaches zero shortly after the scrambling time. 
This early shutoff reflects the property that the collapsing shell becomes effectively invisible to trans-Planckian modes, offering an alternative resolution to the black hole information paradox via a macroscopic remnant.

\end{abstract}
\end{center}

\end{titlepage}

\hide{
\usepackage{amsmath,amssymb,amsfonts}
\allowdisplaybreaks
\usepackage{hyperref}
\usepackage{color}

\def \be {\begin{equation}}
\def \ee {\end{equation}}
\def \bea {\begin{align}}
\def \eea {\end{align}}
\def \nn {\nonumber}
\def \la {\langle}
\def \ra {\rangle}
\def \del {\partial}
\def \Th {\Theta}
\def \rh {\hat{r}}

\setlength{\topmargin}{-4pc}
\setlength{\textheight}{58pc}
\setlength{\textwidth}{6.5in}
\setlength{\oddsidemargin}{-0.2cm}
\setlength{\evensidemargin}{-0.2cm}
\setlength{\footskip}{2.5pc}

\setlength{\fboxrule}{1pt}
\newcommand{\boxedeq}[1]{
  \vskip5pt
  \fcolorbox{blue}{white}{
      \begin{minipage}{5.7in}
      \color{black}
      \be #1\ee
      \vskip2pt
      \end{minipage}
    }
    \vskip10pt
}

\title{HR in SFT: A Conserved Inner Product}
\author{}
\date{}

\begin{document}
\maketitle

\baselineskip 18pt

}


\section{Introduction}
\label{sec:Introduction}


Hawking's prediction~\cite{Hawking:1974rv, Hawking:1974sw} of thermal radiation from a black hole relies on the following feature of Schwarzschild geometry. 
The frequency of an in-mode that becomes an observed Hawking quantum at retarded time $u$ had an initial frequency that is exponentially higher by a factor of $e^{u/2a}$, where $a = 2 G_N M$ is the Schwarzschild radius.
After the scrambling time $u_{scr} \equiv 2a \log(a/\ell_p)$,\footnote{The scrambling time is sometimes defined to differ by a factor of 2 as $2a\log(a^2/\ell_p^2)$. This difference is insignificant as we only need to define the scrambling time as a time scale $\mathcal{O}(a\log(a/\ell_p))$.} where $\ell_p$ is the Planck length, this initial frequency is exponentially trans-Planckian, raising the question of whether the prediction is robust against new physics at the Planck scale.
While generic local UV cutoffs and Lorentz-violating modified-dispersion models leave the standard result about Hawking radiation essentially unchanged~\cite{Unruh:1994je, Corley:1996ar, Unruh:2004zk, Agullo:2009wt}, examples of UV models that lead to suppressed Hawking radiation due to certain spacetime uncertainty relations were found~\cite{Chau:2023zxb,Ho:2023tdq,Ho:2024tby,Ho:2026xhu}.
In particular, it was argued in ref.~\cite{Ho:2023tdq} that the nonlocal modification motivated by string field theory (SFT) --- in which interactions are smeared by the operator $e^{\ell^2\Box}$ --- leads to an early termination of Hawking radiation.

In ref.~\cite{Ho:2023tdq}, Hawking radiation is argued to stop around the scrambling time due to the exponential suppression of interactions in the high-energy limit in string field theory.
The derivation relies on the Hamiltonian formulation proposed there, and later improved in ref.~\cite{Chang:2024scn}.
However, since the Hamiltonian formulation of string field theory is still not fully established, the purpose of this work is to establish this result through a different approach, without referring to the Hamiltonian formulation.

We follow Hawking's original work~\cite{Hawking:1974sw, Hawking:1976ra}, and consider the propagation of waves in the Vaidya background of a collapsing matter shell, and see how positive- and negative-frequency modes in the infinite past are mixed compared with positive- and negative-frequency modes in the future at large distances.
The only difference is that the wave equation is modified in a way to mimic interactions in string field theory.

String field theory is only well understood in a flat background.
In the Lagrangian density, all fields appear in the interaction terms only in the form~\cite{SFT}
\be
\tilde{\phi} \equiv e^{\ell^2\Box} \phi,
\label{phitilde}
\ee
where $\Box = - \del_t^2 + \vec{\del}^2$, where $\ell$ is a parameter of order the string length scale.
In the Wick-rotated momentum space $k$, this factor $e^{\ell^2\Box} \rightarrow e^{- \ell^2 k^2}$ leads to an exponential suppression of the interaction at high energies.

In this work, we incorporate this exponential suppression into the gravitational interaction between the radiation field and the Vaidya background. \footnote{
The effects of such nonlocal form factors have previously been studied in the context of black holes in ref.~\cite{Koshelev:2024wfk}, where it was shown that the interior singularity is resolved.
}
For simplicity of discussion, we shall not distinguish the string length scale $\ell$ from the Planck length scale $\ell_p$.
Due to this exponential suppression, when the momentum $p$ conjugate to $r$ is trans-Planckian ($|p|\ell \gg 1$), the collapsing shell is invisible to the trans-Planckian modes.
Our main result is that the number expectation of an outgoing Gaussian wave packet of width $\Delta u$ centered at retarded time $u_0$ reproduces the standard Planck spectrum at the Hawking temperature $T_H = 1/(4\pi a)$ before the scrambling time ($u_0 \lesssim u_{scr} \equiv 2a \log(a/\ell)$), but is Gaussian-suppressed by a factor $\sim \exp[-2(u_0 - u_{scr})^2/(\Delta u)^2]$ after the scrambling time ($u_0 \gtrsim u_{scr}$).

The rest of the paper is organized as follows.
In Sec.~\ref{sec:action}, we derive the SFT-motivated wave equation for the s-wave reduction of a massless scalar in the Vaidya background.
In Sec.~\ref{sec:outgoing}, we construct the outgoing mode solutions in momentum space and analyze their IR ($|p|\ell \ll 1$) and UV ($|p|\ell \gg 1$) asymptotics.
Then, we compute the Bogoliubov coefficients and obtain the time-dependent Hawking spectrum in Sec.~\ref{sec:hawking}.
We conclude in the final section.
Throughout the paper, we use the mostly-plus signature $\eta_{\mu\nu} = \mathrm{diag}(-,+,+,+)$ and natural units $\hbar = c = 1$.


\section{Massless scalar in the Vaidya background}
\label{sec:action}


The curved spacetime for a dynamical black hole can be described as a gravitational deformation of the Minkowski space.
For instance, the Vaidya metric for a collapsing matter sphere at the speed of light is of the form
\begin{align}
ds^2 &= - \left(1 - \frac{a(v)}{r}\right) dv^2 + 2 dv dr + r^2 (d\th^2 + \sin^2\th d\varphi^2)
\nn \\
&= ds_{\text{Mink}}^2 + \frac{a(v)}{r} dv^2 = g_{\mu\nu} dx^{\mu} dx^{\nu},
\label{eq:Vaidya-metric}
\end{align}
where $ds_{\text{Mink}}^2$ is the Minkowski metric
\be
ds^2_{\rm Mink} = -dv^2 + 2\, dv\, dr + r^2 (d\theta^2 + \sin^2\theta\, d\varphi^2),
\label{Mink-metric}
\ee
and $a(v)$ is defined by
\be
a(v) \equiv 2 G_N M(v)
\ee
with $M(v)$ specifying the mass profile of the collapsing sphere.

For simplicity, we shall assume that the collapsing matter has the profile of a thin shell.
Without loss of generality, we choose the surface of the collapsing shell at $v = 0$, so that 
\begin{align}
a(v) = \left\{
\begin{array}{ll}
a & (v > 0),
\\
0 & (v < - d),
\end{array}
\right.
\label{eq:thin-shell}
\end{align}
where $a$ is the Schwarzschild radius of the black hole, and $d$ is the thickness of the shell. 
The thickness $d$ is assumed to be much smaller than $a$ for the simplicity of deriving Hawking radiation.
The spacetime inside the shell ($v < - d$) is Minkowski space.

The kinetic term of a massless scalar $\phi$ is
\begin{align}
g^{\mu\nu} \del_{\mu}\phi \del_{\nu}\phi =
\eta^{\mu\nu} \del_{\mu}\phi \del_{\nu}\phi - \frac{a(v)}{r} (\del_r \phi)^2.
\end{align}
The correction term $\frac{a(v)}{r} dv^2$ to the Minkowski metric leads to a modification of the Lagrangian as an interaction term $- \frac{a(v)}{r} (\del_r \phi)^2$. 
This allows us to define Hawking radiation as a scattering process in the Minkowski background~\cite{Aoude:2024sve}.

\subsection{SFT-motivated wave equation}

For simplicity, we consider the dimensionally reduced theory of s-waves.
Substituting the s-wave decomposition $\phi^{(4D)}(v, r, \theta, \varphi) = \phi(v, r)/r$ into the four-dimensional action
\be
S^{(4D)} = 
-\frac{1}{2}\int d^4 x\, \sqrt{-g} g^{\mu\nu} \del_\mu \phi^{(4D)} \del_\nu \phi^{(4D)} ,
\ee
one obtains the action for the (rescaled) radial mode $\phi(v, r)$:
\begin{align}
S = - \frac{1}{2} \int dv dr \left(2 \del_v\phi \del_r \phi + (\del_r\phi)^2
- \frac{a(v)}{r} (\del_r \phi)^2 + V(v, r) \phi^2\right)
\end{align}
up to boundary terms and an overall factor of $4\pi$ absorbed in the field normalization.
The potential
\be
V(v, r) \equiv \frac{a(v)}{r^3}
\ee
affects only the grey-body factors and is subleading at large $r$, so we drop it in what follows.
Hawking radiation and the Hawking temperature are insensitive to $V(v, r)$~\cite{Page:1976df, Hawking:1976de}.
Ignoring the potential term, the action is
\begin{align}
S = - \frac{1}{2} \int dv dr \left(2 \del_v\phi \del_r \phi + (\del_r\phi)^2
- \frac{a(v)}{r} (\del_r \phi)^2\right).
\label{action}
\end{align}
This is the standard s-wave simplification.

Interpreting the background geometry as a coherent state of the gravitational field in the Minkowski spacetime, the interaction term $-(a(v)/r)(\del_r \phi)^2$ in eq.~\eqref{action} can be read as a three-point vertex with the graviton in a coherent state, and the prefactor $a(v)/r$ is its expectation value.
In string field theory, every quantum-field leg of an interaction vertex carries the smearing factor $e^{\ell^2\Box}$.
Each $\phi$ entering the vertex is replaced by $\tilde\phi$ defined in eq.~\eqref{phitilde}.\footnote{
For simplicity, we choose the potential $a(v)/r$ to be not smeared by the factor $e^{\ell^2\Box}$.
Smearing $a/r$ additionally suppresses the high-$p$ vertex, so it can only strengthen the invisibility of the shell to the high-energy modes, and further enhance the termination of Hawking radiation that we will establish below.
}
Thus, to mimic the SFT effect, the interaction term should be modified as
\begin{align}
- \frac{a(v)}{r} (\del_r \tilde{\phi})^2,
\end{align}
where $\tilde{\phi}$ is defined in eq.~\eqref{phitilde} and
\be
\Box \equiv \eta^{\mu\nu} \del_{\mu} \del_{\nu} = \del_r(2\del_v + \del_r)
\label{Box}
\ee
is the two-dimensional d'Alembertian in $(v, r)$ coordinates.

The action~\eqref{action} is now
\begin{align}
S = - \frac{1}{2} \int dv dr \left(2 \del_v\phi \del_r \phi + (\del_r\phi)^2
- \frac{a(v)}{r} (\del_r \tilde{\phi})^2\right) .
\label{action-SFT}
\end{align}
By applying the variational least-action principle to the nonlocal action, we arrive at the full equation of motion
\be
\del_r\left((2\del_v + \del_r) \phi - e^{\ell^2\Box} \frac{a(v)}{r} \del_r e^{\ell^2\Box} \phi\right) = 0.
\label{wave-eq}
\ee
Isolating the ingoing and outgoing components, the SFT-motivated nonlocal wave equation for the radial outgoing modes takes the explicit form
\be
(2\del_v+\del_r) \phi(v,r) - e^{\ell^2\Box} \, \frac{a(v)}{r} \del_r \, e^{\ell^2\Box} \, \phi(v,r)\;=\;0 ,
\label{eq:eom-phi}
\ee
which is equivalent to
\begin{align}
\left(r(2\del_v + \del_r) e^{-\ell^2\Box} - a(v) \del_r e^{\ell^2\Box} \right)\phi = 0 ,
\label{wave-eq-SFT}
\end{align}
where $\Box$ is given by eq.~\eqref{Box}.

Expressed in the momentum space with
\be
\phi(v,r)=\int\!\frac{dp}{\sqrt{2\pi}}\,\Phi(v,p)\,e^{ipr} ,
\ee
the outgoing wave equation~\eqref{wave-eq-SFT} is equivalent to
\begin{align}
\left[
i\del_p(2\del_v + ip) e^{\ell^2(p^2 - 2i p\del_v)} - i a(v) p e^{- \ell^2 (p^2 - 2 i p\del_v)}
\right] \Phi(v, p) = 0.
\label{wave-eq-SFT-momentum}
\end{align}

Since eqs.~\eqref{wave-eq-SFT} and~\eqref{wave-eq-SFT-momentum} involve $v$-derivatives of an infinite order, standard notions such as the Hamiltonian and Hilbert space do not necessarily apply.
An attempt at the Hamiltonian formulation of such a quantum system can be found in ref.~\cite{Chang:2024scn}.
But we will not rely on the Hamiltonian formulation in this work.

We shall solve eq.~\eqref{wave-eq-SFT-momentum} below to derive Hawking radiation.
Before that, it is already clear from the dependence on the exponential factors $e^{\pm \ell^2 p^2}$ of the two terms that the second term is exponentially suppressed for trans-Planckian modes with $|p| \gg 1/\ell$.
This reflects the soft UV limit of string interactions, and implies that the solutions of trans-Planckian modes are insensitive to the collapsing profile $a(v)$.
The collapsing matter is in practice invisible to the high-energy modes in the limit $|p| \rightarrow \infty$.

\subsection{Inner product}
\label{sec:innerproduct}

We will need a conserved inner product to derive Hawking radiation.
However, due to the infinite $v$-derivatives, the usual inner product\footnote{In 2D, the coordinate $r$ lives in $\mathbb{R}$.} for a Klein-Gordon field,
\begin{align}
\la \phi_1, \phi_2 \ra(v) &\equiv
- i \int_{-\infty}^{\infty} dr \Big[
\phi_1^{*}(v,r) \del_r\phi_2(v,r) - \bigl(\del_r\phi_1^{*}(v,r)\bigr) \phi_2(v,r)
\Big]
\nn \\
&=
2 \int_{-\infty}^{\infty} dp \, p \, \Phi_1^{*}(v,p)\Phi_2(v,p) ,
\label{eq:innerproduct}
\end{align}
is no longer conserved for the wave equation~\eqref{wave-eq-SFT}.
On the other hand, at least formally, a conserved inner product exists.
Complexifying $\phi$, the action~\eqref{action-SFT} is invariant under the U(1) phase rotation $\phi \to e^{i\alpha} \phi$, and Noether's theorem ensures the existence of a current $j^{\mu}_c$ that is conserved on-shell: $\del_{\mu} j^{\mu}_c = 0$~\cite{Llosa:1993sj, HerediaPimienta:2023ogb}.
As the complexified field and the real field satisfy the same wave equation for a quadratic Lagrangian, the corresponding conserved charge
\be
\la \phi_1, \phi_2 \ra_c (v) \equiv \int_{-\infty}^{\infty} dr \, j^v_c[\phi_1, \phi_2]
\label{eq:jc}
\ee
can be used as the conserved inner product for the SFT-modified dynamics.

Because the nonlocal modification of the action~\eqref{action-SFT} enters only through the term $- \frac{a(v)}{r} (\del_r \tilde\phi)^2$, the current decomposes as
\be
j^{\mu}_c[\phi_1, \phi_2] = j^{\mu}_{\rm KG}[\phi_1, \phi_2] + \frac{a(v)}{r} \Delta j^{\mu}[\phi_1, \phi_2; \ell^2] ,
\label{eq:current-decomp}
\ee
where $j^{\mu}_{\rm KG}$ is the standard Klein--Gordon current generating~\eqref{eq:innerproduct}, and $\Delta j^{\mu}$ is bilinear in $\phi_1, \phi_2$.
For the thin-shell configuration~\eqref{eq:thin-shell} under consideration, the correction $\Delta j^{\mu}$ vanishes identically inside the collapsing shell where $a(v) = 0$, so that the conserved inner product agrees with the usual inner product, i.e., $\la \cdot, \cdot \ra_c = \la \cdot, \cdot \ra$, on any slice inside the collapsing shell.
Similarly, in the asymptotically flat region ($r \to \infty$), the correction is again suppressed by the factor $1/r$, so that $\la \cdot, \cdot \ra_c \simeq \la \cdot, \cdot \ra$.

The key technical ingredient needed for our derivation of Hawking radiation is the conserved inner product $\la \cdot, \cdot \ra_c$~\eqref{eq:jc} that is approximated by the ordinary Klein--Gordon inner product $\la \cdot, \cdot \ra$~\eqref{eq:innerproduct} in both asymptotic regions $v \to - \infty$ and $r \to \infty$.
Although the $U(1)$ Noether current~\eqref{eq:current-decomp} is constructed formally as an infinite series in $\ell^2$, and its convergence is not guaranteed, all we need is the existence of a conserved inner product and its approximation by the standard Klein-Gordon inner product in the (asymptotically) flat regions.
As long as the notion of inner product cannot be defined in a theory with the same nonlocality as string field theory, our derivation does not depend on the details of the conserved inner product.

\section{Outgoing wave solution}
\label{sec:outgoing}

In preparation for deriving Hawking radiation, we solve the outgoing wave equation~\eqref{eq:eom-phi} in this section.
First, for a fixed frequency $\omega$, we look for outgoing solutions in the region outside the shell ($v>0$) of the form
\be
\phi^{\rm out}_{\om}(v,r) = N_{\om} e^{-i\omega v}\!\int_{-\infty}^{\infty}\!\frac{dp}{\sqrt{2\pi}}\;e^{ipr}\,e^{F_\omega(p)} .
\label{eq:ansatz}
\ee


\subsection{Solution outside the collapsing shell}

For $v>0$, eq.~\eqref{eq:eom-phi} reduces to
\be
\hat{r} e^{\ell^2 p(p-2\omega)} (p-2\om) e^{F_\omega(p)} = a p e^{- \ell^2 p(p-2\omega)}e^{F_\omega(p)} ,
\label{eq:integraleq}
\ee
\noindent
where $\hat{r} = i \del_p$, for a single-frequency mode~\eqref{eq:ansatz}.
Eq.~\eqref{eq:integraleq} gives a differential equation for $F_{\om}(p)$:
\be
i e^{\ell^2 p(p-2\om)} \left[1 + 2\ell^2 (p-\om)(p-2\om) + (p-2\om)\del_p F_{\om}(p)\right]
= a p e^{- \ell^2 p(p-2\om)} .
\ee
It is a first-order ODE that can be easily solved by
\be
F_\omega(p) = - \ell^{2}(p-\omega)^{2} - \log(p-2\om-i\eps) 
- ia \int_0^{p} dp' \, \frac{p' \, e^{-2\ell^{2}p'(p'-2\omega)}}{p'-2\om-i\eps} + \mathrm{const},
\label{eq:Fsolution}
\ee
where Feynman's $i\eps$ prescription is applied and the constant may depend on $\om$ and $a$ but not $p$.

\subsection{UV and IR regimes of $F_\omega(- p)$}
\label{sec:limits}

\hide{
We rewrite eq.~\eqref{eq:Fsolution} as
\begin{align}
F_\omega(p) &=
- \ell^{2}(p-\omega)^{2} 
- \log(p-2\om-i\eps) 
- ia \int_0^{p} dp' \frac{p' e^{-2\ell^{2}p'(p'-2\omega)}}{p'-2\om-i\eps} 
+ {\rm const} .
\label{eq:exact-F}
\end{align}
where we add and subtract the $\ell = 0$ form of the integral, so that the second and third terms reproduce the standard Hawking IR result, and the first and fourth carry all the nonlocal corrections.
}

For $|p| \ll 1/\ell$, the solution~\eqref{eq:Fsolution} can be approximated by
\begin{align}
e^{F_{\om}(p)} &\simeq
{\rm const} \times 
e^{- i a p} (p - 2\om - i\eps)^{- (1 + 2ia\om)} .
\label{eq:eF-low}
\end{align}
This is the standard result in the low-energy effective theory.

For $|p| \gg 1/\ell$, the 3rd term on the right-hand side of eq.~\eqref{eq:Fsolution} is exponentially suppressed.
The first two terms in eq.~\eqref{eq:Fsolution} are independent of $a$.
Effectively, the collapsing shell is absent from the viewpoint of the trans-Planckian modes.
The approximate solution is
\begin{align}
e^{F_{\om}(p)} &\simeq
{\rm const} \times
e^{- \ell^2 (p - \om)^2} e^{- i {\rm sgn}(p) \sqrt{\frac{\pi}{8}} \, a/\ell} e^{-2ia\om\log(a/\ell)} (p - 2\om - i\eps)^{- 1} .
\label{eq:eF-high}
\end{align}

For simplicity, we shall approximate the exact solution~\eqref{eq:Fsolution} by eq.~\eqref{eq:eF-low} for $|p| \leq 1/\ell$, and by eq.~\eqref{eq:eF-high} for $|p| > 1/\ell$.

The only intrinsic $a$-dependence in eq.~\eqref{eq:eF-high} is the $\om$- and $p$-independent phase $e^{\pm i\sqrt{\pi/8} \, a/\ell}$, which can be absorbed in the arbitrary constant factor and has no effect on Hawking radiation. 
In this sense, the collapsing shell is invisible to the trans-Planckian modes, and this result is expected to be unchanged when we include the region $v \in (-d, 0)$ where $a(v)$ is not a constant.
This fact should have already been clear by merely looking at the differential equation~\eqref{wave-eq-SFT-momentum}.


\subsubsection{The IR regime: $|p|\,\ell\ll 1$}

The solution $F_{\om}(p)$ will be needed in the calculation of Hawking radiation for $p < 0$.
That is, we are interested in $F_{\om}(- p)$ for $p > 0$.

For $\ell|p|\ll 1$, choosing the overall constant factor to be $1/a$ so that $e^{F_{\om}}$ is dimensionless, we have
\be
e^{F_{\om}(-p)} \simeq - \frac{1}{a} 
e^{iap} (p+2\om)^{-(1+2ia\om)} e^{- 2\pi a\om} .
\label{eq:eF-low-p<0}
\ee
Examining the outgoing mode solution at large $r$ in the asymptotically flat region, we determine the normalization constant in eq.~\eqref{eq:ansatz} as
\be
|N_{\om}|^{2} \equiv \frac{a^3}{2\pi(1-e^{-4\pi a\omega})} ,
\label{eq:N0}
\ee
so that $\phi^{\rm out}_{\om}(v, r)$~\eqref{eq:ansatz} is a normalized outgoing mode at low energies ($\om \ll \ell^{-1}$).
Note that this factor~\eqref{eq:N0} is the standard result completely determined by the normalization of the low-energy modes.

Since the nonlocal operator $e^{\ell^2\Box}$ reduces to the identity at long wavelengths, the standard continuity of $\phi$ at $v=0$ can be applied to determine the wave function inside the shell.
We will see in the next section that this reproduces the standard Hawking radiation before the scrambling time.

\subsubsection{The UV regime: $|p|\,\ell\gg 1$}

Again, we introduce an overall factor of $1/a$ for $e^{F_{\om}}$ to be dimensionless.
Then, one can approximate $e^{F_{\om}(-p)}$ for $|p| \ell \gg 1$ by
\be
e^{F_\omega(- p)} \simeq 
- \frac{1}{a} e^{i\sqrt{\pi/8} \, a/\ell} e^{-2ia\om\log(a/\ell)} \frac{e^{-\ell^{2}(p+\omega)^{2}}}{p+2\omega} e^{-2\pi a\om} ,
\label{eq:psi-UV-2}
\ee
Apart from the constant phase $e^{i\sqrt{\pi/8} \, a/\ell}$ already present in eq.~\eqref{eq:eF-high}, there are two further $a$-dependent factors in eq.~\eqref{eq:psi-UV-2}: the phase $e^{-2ia\om\log(a/\ell)}$ and the Boltzmann factor $e^{-2\pi a\om}$.
The latter is inherited from the branch-cut crossing of $\log(p-2\om-i\eps)$ at $p=2\om$ in the IR. 
Neither is generated by UV physics.
In the calculation of Hawking radiation below, the phase $e^{-2ia\om\log(a/\ell)}$ will be responsible for the cutoff of the Hawking wave packet at the scrambling time. 
The trans-Planckian modes, therefore, contribute to Hawking radiation only through inherited IR structure.
The collapsing matter (the black hole) is essentially invisible to the trans-Planckian modes, as we have commented before and after eq.~\eqref{eq:eF-high}.
This feature is expected not to change for a generic profile $a(v)$ of the collapsing matter.

\section{Hawking radiation}
\label{sec:hawking}

In this section, we calculate the vacuum expectation value of the number operator $\la \hat{N}_\Psi\ra$ for Hawking particles detected at any retarded time $u$ as a characterization of the time-dependent Hawking radiation.
We shall see that, due to the exponential suppression of trans-Planckian interactions, Hawking radiation terminates around the scrambling time.

\subsection{Wave packet and number operator}
\label{sec:chain}

We compute the vacuum expectation value of the number operator for a Hawking quantum in a Gaussian wave packet centered at a certain retarded time $u = u_0$:
\begin{align}
\Phi_\Psi(v,r) &\equiv
\int_0^\infty\!d\omega\,\Psi(\omega)\,\phi^{\rm out}_\omega(v,r) ,
\label{eq:Phi-Psi}
\end{align}
where
\be
\Psi(\omega)\;=\;(2\pi\sigma^{2})^{-1/4}\;e^{-(\omega-\omega_0)^{2}/(4\sigma^{2})}\;e^{i\omega u_0} .
\label{eq:Psi}
\ee
It has a central frequency $\omega_0 \sim 1/a$ and width $\sigma \ll \om_0$. 
The normalization condition on the wave packet is
\be
\int_0^\infty\!|\Psi(\omega)|^{2}\,d\omega \simeq 1 .
\ee

The annihilation operator outside the shell can be defined as $\hat{b}_\omega=\la\phi^{\rm out}_\om, \hat{\phi}\ra$ in terms of the inner product~\eqref{eq:innerproduct}.
At large $r$ and large $v$, in the asymptotically flat region, we expand the field $\hat\phi$ in positive and negative frequency modes:
\be
\hat{\phi}(v,r)
=
\int_0^\infty \frac{d\om}{\sqrt{4\pi\om}}
\bigl[ \hat{b}_{\om} e^{- i\om (v - 2r)} + \hat{b}^{\dagger}_{\om} e^{i\om (v - 2r)} \bigr].
\label{eq:phi-out}
\ee
We define the annihilation operator for a given wave packet $\Psi$ as
\begin{align}
\hat{b}_\Psi &\equiv
\int_0^\infty\!d\omega\,\Psi^{*}(\omega) \hat{b}_\om
= \la\Phi_\Psi, \hat{\phi}\ra ,
\label{eq:bPsi-inner}
\end{align}
where $\la \cdot, \cdot \ra$ denotes the usual inner product for a Klein--Gordon scalar field.

According to our discussions on the inner product in Sec.~\ref{sec:innerproduct},
\be
\hat{b}_\Psi = \la\Phi_\Psi, \hat{\phi}\ra\Big|_{v \rightarrow \infty}
\simeq \la\Phi_\Psi, \hat{\phi}\ra_c\Big|_{v \rightarrow \infty}
=  \la\Phi_\Psi, \hat{\phi}\ra_c\Big|_{v \rightarrow - \infty}
= \la\Phi_\Psi, \hat{\phi}\ra\Big|_{v \rightarrow - \infty} ,
\label{eq:bPsi-at-zero}
\ee
where the second equality is on-shell conservation of $\la \cdot, \cdot \ra_c$, 
\footnote{
This approximation is justified because, in the asymptotic future $v\to\infty$, the wave packet $\Phi_{\Psi}$ is localized in the region $r\to\infty$, where the two inner products coincide (see eq.~\eqref{eq:current-decomp}).
}
the third is exact (the slice lies inside the shell, where $a(v) = 0$), and only the first equality involves an approximation.

Our goal is to compute the number of Hawking particles of this wave packet:
\be
\la \hat{N}_\Psi\ra \equiv \la 0| \hat{b}^{\dagger}_\Psi \hat{b}_\Psi |0\ra .
\ee
The vacuum $|0\ra$ is by definition annihilated by the annihilation operators $\hat{a}_{\omega'}$ inside the collapsing shell. 
That is, we choose the initial state to be the Minkowski vacuum in the infinite past.
We shall evaluate the expectation value of the number operator $\langle\hat{N}_\Psi\rangle$ for the outgoing Gaussian wave packet by mapping the late-time annihilation operators to the early-time Unruh vacuum $|0\ra$. 
The Hawking spectrum is computed via the integrated Bogoliubov coefficient.

From eq.~\eqref{eq:ansatz}, eq.~\eqref{eq:Phi-Psi} inside the collapsing shell at an arbitrary point $v=v_0 < -d$ (one can take $v_0 \rightarrow - \infty$, but the result is independent of $v_0$ as long as it is inside the collapsing shell) can be evaluated as
\begin{align}
\Phi_\Psi(v_0,r)\;=\;\int_{-\infty}^{\infty}\!\frac{dp}{\sqrt{2\pi}}\,e^{ipr}\,A_\Psi(p),
\label{eq:Phi-Psi-A}
\end{align}
where
\be
A_\Psi(p) \equiv \int_0^\infty\!d\omega \Psi(\omega) N_\om e^{F_\omega(p)} e^{-i\om v_0} .
\label{eq:Adef}
\ee

On the $v=v_0$ slice inside the shell, we have
\be
\phi(v_0,r) = \int_0^\infty\!\frac{d\omega'}{\sqrt{4\pi\omega'}} 
\bigl(\hat{a}_{\om'} e^{- i\omega' (v_0 - 2r)} + \hat{a}^{\dagger}_{\om'} e^{i\omega' (v_0 - 2r)} \bigr),
\label{eq:phi-in}
\ee
where $\hat{a}_{\om}$ and $\hat{a}^{\dag}_{\om}$ are the annihilation and creation operators for a free field in the Minkowski spacetime.
Plugging eqs.~\eqref{eq:Adef} and \eqref{eq:phi-in} into $\hat{b}_\Psi=\la\Phi_\Psi,\phi\ra$, we find
\be
\hat{b}_\Psi = 
\int_0^\infty\!d\omega' \bigl(\alpha_\Psi^{*}(\omega') \hat{a}_{\omega'} 
- \beta_\Psi^{*}(\omega') \hat{a}^{\dagger}_{\omega'}\bigr) ,
\label{eq:bPsi-in}
\ee
with
\be
\alpha_\Psi(\omega') = 2\sqrt{2\omega'} A_\Psi(2\omega') e^{i\omega' v_0} , 
\qquad
\beta_\Psi(\omega') = 2\sqrt{2\omega'} A_\Psi(-2\omega') e^{- i\omega' v_0} .
\ee

Using $\la 0|\hat{a}_{\omega''}\hat{a}^{\dagger}_{\omega'}|0\ra=\delta(\omega''-\omega')$, we find
\begin{align}
\la \hat{N}_\Psi\ra &= 
\int_0^\infty d\omega' |\beta_\Psi(\omega')|^{2}
= 2 \int_0^\infty dp p |A_\Psi(-p)|^{2} ,
\label{eq:N-omegapr}
\end{align}
where
\begin{align}
A_\Psi(-p) = \int_0^\infty d\om \Psi(\omega) N_\om e^{F_\omega(-p)} 
\label{eq:master}
\end{align}
up to an irrelevant phase that we shall omit from now on.

\subsection{Approximation}

We split \eqref{eq:master} into IR and UV contributions according to the approximate solution of $F_\omega(p)$ studied in \S\ref{sec:limits}:
\be
\la \hat{N}_\Psi\ra
= I_{\rm IR} + I_{\rm UV} ,
\label{eq:split}
\ee
where the UV and IR contributions are
\begin{align}
I_{\rm IR} &\equiv
2\!\int_0^{1/\ell}\!dp\;p\;|A_\Psi(-p)|^{2} ,
\\
I_{\rm UV} &\equiv
2\!\int_{1/\ell}^{\infty}\!dp\;p\;|A_\Psi(-p)|^{2} .
\end{align}

\subsubsection{IR contribution}

Substituting the IR approximation~\eqref{eq:eF-low-p<0} into $A_\Psi(-p)$, we find
\be
A_\Psi(-p) \simeq 
({\rm phase})\times
\int_0^\infty\!d\omega\,\Psi(\omega) N_\omega e^{-2\pi a\omega} (a(p+2\omega))^{-(1+2ia\omega)}
\ee
up to an overall constant phase factor.
Due to the Gaussian wave packet $\Psi$~\eqref{eq:Psi}, all slowly varying functions of $\om$ (e.g., $\log(a(p+2\om))$) can be approximated by setting $\om = \omega_0$ (it produces only $O(\sigma)$ subleading shifts). 
We can thus approximate $|A_\Psi(-p)|^{2}$ as
\be
|A_\Psi(-p)|^{2} \simeq
2\sqrt{2\pi} \sigma |N_{\om_0}|^{2} \frac{e^{-4\pi a\omega_0}}{a^2(p+2\omega_0)^{2}} e^{-2\sigma^{2}T(p)^{2}} ,
\label{eq:Aip-IR}
\ee
where
\begin{align}
T(p) \equiv u_0\,-\,2a\,\log(a(p+2\omega_0)) .
\label{eq:Tdef-new}
\end{align}
The Gaussian integral is dominated by the integration around $p_{\star}$ determined by
\be
T(p_{\star}) = 0 \quad\Longleftrightarrow\quad 
p_{\star}(u_0) = \frac{1}{a} e^{u_0/(2a)}-2\omega_0 
\label{eq:saddle}
\ee
The $p$-integral is dominated by the IR domain at early times (small $u_0$), and by the UV domain at late times (large $u_0$).

Changing variable $T=u_0-2a\log(a(p+2\om_0))$, and taking $p\gg 2\omega_0$ (due to the factor $e^{-2\sigma^{2}T(p)^{2}}$), the IR integral becomes
\be
I_{\rm IR}\;\simeq\;\frac{4\sqrt{2\pi}\,\sigma\,|N_{\om_0}|^{2}\,e^{-4\pi a\omega_0}}{2a^3}\!\int_{u_0-u_{scr}}^{u_0-2a\log(2a\om_0)}\!dT\,e^{-2\sigma^{2}T^{2}} ,
\ee
with the scrambling time $u_{scr}$ defined by
\be
u_{scr} \equiv 2a\log(a/\ell) .
\label{eq:uscr}
\ee

Let us now discuss the situations separately when $u_0 \lesssim u_{scr}$ and $u_0 \gtrsim u_{scr}$.
Hawking radiation starts to appear around $u \sim 2a \log(2a\om_0)$, so we will always assume $u \gg 2a\log(2a\om_0)$.

\paragraph{(i) $u_0 \lesssim u_{scr}$:} 
The Gaussian integrates to $\sqrt{\pi/(2\sigma^{2})}$, giving
\be
I_{\rm IR}\;\simeq\;\frac{2\pi}{a^3}\,|N_{\om_0}|^{2}\,e^{-4\pi a\omega_0} ,
\ee
where $|N_{\om_0}|^2$ is given by eq.~\eqref{eq:N0}.
We get the Planck distribution:
\be
I_{\rm IR} \simeq
\frac{1}{e^{4\pi a\omega_0}-1} \qquad (u_0 \lesssim u_{scr}) .
\label{eq:Hawking}
\ee
This is the universal Hawking spectrum at temperature $T_H=1/(4\pi a)$. 
It is insensitive to the parameters $u_0$ (as long as it lies in the range $(2a\log(2a\om_0), u_{scr})$), $\sigma$, and the wave-packet shape.

\paragraph{(ii) $u_0\gtrsim u_{scr}$:}
The lower limit $(u_0-u_{scr})$ of integration is now positive and large.
The remaining Gaussian integral is a complementary error function,
\begin{align}
I_{\rm IR}
&\simeq\;
\frac{1}{2} \frac{1}{e^{4\pi a\omega_0}-1} \mathrm{erfc}\bigl(\sigma \sqrt{2}(u_0-u_{scr})\bigr)
\nn \\
&\simeq
\frac{1}{2\sqrt{2\pi}} \frac{1}{e^{4\pi a\omega_0}-1} \frac{e^{- 2 \sigma^2 (u_0-u_{scr})^2}}{\sigma (u_0-u_{scr})}
\qquad (\sigma (u_0 - u_{scr}) \gg 1).
\label{eq:IIR-large-u}
\end{align}
which goes to $0$ exponentially fast after the scrambling time.
This essentially reproduces the conclusion of ref.~\cite{Ho:2022gpg}, namely that Hawking radiation terminates at approximately the scrambling time when a UV cutoff $|p|\leq 1/\ell$ is imposed.

The cross-over at $u_0 \sim u_{scr}$ is governed by the complementary error function
\be
I_{\rm IR}(u_0) = \frac{1}{2} \frac{1}{e^{4\pi a\om_0}-1} \mathrm{erfc}(\sqrt{2}\sigma(u_0 - u_{scr})) 
\ee
which interpolates eq.~\eqref{eq:Hawking} and eq.~\eqref{eq:IIR-large-u}.
The transition window $|u_0 - u_{scr}| \lesssim 1/\sigma$ is set by the spectrum width $\sigma \ll \om_0$ of the wave packet.

\subsubsection{UV contribution}

In the UV regime, we have eq.~\eqref{eq:psi-UV-2}, and thus eq.~\eqref{eq:master} gives
\begin{align}
A_\Psi(-p) &\simeq
-\,\frac{N_{\omega_0}\, e^{i\sqrt{\pi/8} \, a/\ell} e^{-\ell^{2}(p+\om_0)^{2}}}{a(p+2\omega_0)}\,\hat\Psi(u_{scr}) e^{-2\pi a\om_0} ,
\end{align}
where $u_{scr}$ is defined by eq.~\eqref{eq:uscr}, and
\begin{align}
\hat\Psi(u_{scr})
&\equiv
\int_0^\infty\!d\omega\,\Psi(\omega) e^{-i\om u_{scr}}.
\end{align}
The Gaussian integral evaluates to
\be
|\hat\Psi(u_{scr})|^{2} \simeq 2\sqrt{2\pi}\,\sigma\,e^{-2\sigma^{2}(u_0 - u_{scr})^{2}} ,
\ee
so that
\be
|A_\Psi(-p)|^{2}
\simeq
\frac{|N_{\omega_0}|^{2} e^{-2\ell^{2}(p+\om_0)^{2}}}{a^2(p+2\om_0)^{2}}\times 2\sqrt{2\pi} \sigma e^{-2\sigma^{2}(u_0 - u_{scr})^{2}} e^{-4\pi a\om_0} ,
\ee
and therefore
\begin{align}
I_{\rm UV}
&\simeq
4\sqrt{2\pi} \sigma |N_{\omega_0}|^{2} e^{-4\pi a\om_0} 
\int_{1/\ell}^{\infty} \frac{p\,dp}{a^2(p+2\om_0)^{2}} 
e^{-2\ell^{2}(p+\om_0)^{2}} e^{-2\sigma^{2}(u_0 - u_{scr})^{2}}
\nn \\
&\simeq
4\sqrt{2\pi} \sigma |N_{\omega_0}|^{2} e^{-4\pi a\om_0} e^{-2\sigma^{2}(u_0 - u_{scr})^{2}}
\int_{1}^{\infty} \frac{d\bar{p}}{a^2 \bar{p}} e^{-2\bar{p}^{2}}
\nn \\
&=
\frac{2\sqrt{2\pi} \Gamma(0,2) \sigma |N_{\omega_0}|^{2}  e^{-4\pi a\om_0}}{a^2} e^{-2\sigma^{2}(u_0 - u_{scr})^{2}}
\nn \\
&=
\frac{2 \Gamma(0,2)}{\sqrt{2\pi}} \sigma a \frac{1}{e^{4\pi a\omega_0}-1}
e^{-2\sigma^{2}(u_0 - u_{scr})^{2}},
\label{eq:IUV}
\end{align}
where $\Gamma(0,2) \simeq 0.05$ is the incomplete Gamma function.
Here, we identify the Planck distribution $\frac{1}{e^{4\pi a\omega_0}-1}$ but also an exponentially decaying factor $e^{-2\sigma^{2}(u_0 - u_{scr})^{2}}$ after the scrambling time ($u_0 > u_{scr}$).

\subsection{Time-dependence of Hawking radiation}

Putting both IR and UV contributions together, the vacuum expectation value of the number of Hawking particles is given by
\begin{align}
\la \hat{N}_\Psi\ra
&\simeq
\frac{1}{e^{4\pi a\omega_0}-1} \left[
1 + \frac{2 \Gamma(0,2)}{\sqrt{2\pi}} \sigma a
e^{-2\sigma^{2}(u_0 - u_{scr})^{2}}
\right] 
\simeq
\frac{1}{e^{4\pi a\omega_0}-1} 
\qquad (u_{scr} \gg u_0 \gg 2a \log(2a\om_0))
\end{align}
before the scrambling time, and by
\begin{align}
\la \hat{N}_\Psi\ra
&\simeq
\frac{1}{2\sqrt{2\pi}}
\frac{1}{e^{4\pi a\om_0}-1} 
\left[
\frac{1}{\sigma (u_0-u_{scr})}
+ 4 \Gamma(0,2) \sigma a
\right]
e^{- 2 \sigma^2 (u_0 - u_{scr})^2} 
\nn \\
&\simeq 
\frac{2 \Gamma(0,2) \sigma a}{\sqrt{2\pi}}
\frac{1}{e^{4\pi a\om_0}-1}
e^{- 2 \sigma^2 (u_0 - u_{scr})^2}
\qquad \left(u_0 - u_{scr} \gg \frac{1}{4\Gamma(0,2)\sigma^2 a} \sim \mathcal{O}(a)\right)
\label{eq:final}
\end{align}
after the scrambling time.
The radiation is dominated by the IR modes for $u_0 < u_{scr}$ and by the UV modes for $u_0 > u_{scr}$.

In the limit of $\s \rightarrow 0$, Hawking radiation persists.
However, this corresponds to the detection of a Hawking particle at an exact frequency, which, in principle, takes an infinite amount of time.
The uncertainty $\Delta u \sim 1/\s$ of the termination time of Hawking radiation is a consequence of the uncertainty relation $\Delta E \Delta T \gtrsim 1$ for a Gaussian wave packet.

The IR contribution reproduces the usual Hawking spectrum as long as the geometric-optics saddle $p_{\star}(u_0) \simeq (1/a)e^{u_0/(2a)}$ stays in the IR window $p_{\star}\,\ell\ll 1$, i.e., before the scrambling time ($u_0\lesssim u_{scr}\equiv2a\log(a/\ell)$). 
Beyond the scrambling time, the saddle leaves the IR domain through the cutoff at $p=1/\ell$, and $\la \hat{N}_\Psi\ra$ acquires a Gaussian-tail suppression of width $1/\sigma$ in $u_0$. 
The cutoff time $u_{scr} \equiv 2a\log(a/\ell)$ is the moment at which the outgoing quantum being redshifted into an observed Hawking quantum has its initial frequency equal to the string scale inside the shell.

The total energy loss through Hawking radiation before the scrambling time,
\be
\frac{\Delta M}{M} \sim \frac{\ell^2\log(a/\ell)}{a^2} ,
\ee
is a negligible fraction of the initial mass of the black hole.
It is thus safe to ignore the back-reaction of the modified Hawking flux on the geometry as we did in the paper.

\section{Conclusion}
\label{sec:Conclusion}

We have derived the Hawking radiation in the Vaidya background in a hypothetical UV completion motivated by string field theory.
The calculation reproduces the standard Planck spectrum at the Hawking temperature $T_H = 1/(4\pi a)$ for retarded times $u_0 \lesssim u_{scr} \equiv 2a\log(a/\ell)$, but implies a shutoff of Hawking radiation beyond the scrambling time.
The result is independent of the profile $a(v)$ of the collapse, because the collapsing shell is invisible to the trans-Planckian modes that seed the late-time Hawking quanta.

The same scrambling-time shutoff in SFT was first suggested in ref.~\cite{Ho:2023tdq} using a Hamiltonian formulation of SFT proposed there, and further developed in ref.~\cite{Chang:2024scn}. 
Its implications for the black-hole information paradox were explained in ref.~\cite{Ho:2024tby}.
However, the Hamiltonian formulation has not been rigorously established for infinite-derivative Lagrangians, so the conclusion was not rigorous. 
The present derivation circumvents this difficulty.
We work directly with the wave equation to calculate the expectation value of the number operator in the Unruh vacuum.
The principal progress over our earlier publications is therefore methodological.

Most contemporary resolutions of the paradox accept Hawking's semiclassical spectrum --- thermal radiation lasting a black-hole lifetime $u_{\rm Page} \sim a^3/G_N$~\cite{Page:1993wv}. 
New ingredients are added to restore unitarity at the Page time, typically through subtle correlations among the emitted quanta. 
The island formula and the replica-wormhole derivation of the Page curve~\cite{AdS-Entropy}, the AMPS firewall argument~\cite{firewall}, black-hole complementarity~\cite{Susskind:1993if, Lowe:1995ac, Polchinski:1995ta}, the soft-hair proposal~\cite{Hawking:2016msc}, and the final-state proposal of ref.~\cite{Horowitz:2003he} all belong to this category.

The proposal here is different. 
The quantum fluctuations responsible for late-time Hawking quanta are exponentially blueshifted past the string energy scale before reaching the collapsing shell. 
In any local UV completion that preserves the blueshift, the Hawking spectrum is essentially unchanged.
But in the nonlocal SFT completion considered here, the smearing factor $e^{\ell^2\Box}$ suppresses the scattering of trans-Planckian modes off the gravitational background of the shell. 
The radiation therefore stops at $u \sim u_{scr} = 2a\log(a/\ell)$, significantly earlier than the Page time.
Once it stops, the information remains in the stable, large remnant, rather than being recovered through subtle correlations among emitted quanta. 
No replica wormholes, no firewalls, no soft hair, and no fine-tuned final state are required.

A standard objection to remnant scenarios is the infinite-degeneracy problem~\cite{Susskind:1995da,Giddings:1994qt,Chen:2014jwq}.
If a black hole of arbitrary mass evaporates down to a Planck-scale remnant, the resulting remnant species must carry unboundedly many internal states, in conflict with effective field theory expectations. 
The present scenario evades this objection because radiation stops at $u_{scr}$, when the black hole still carries all of its original Bekenstein--Hawking entropy $S_{\rm BH} \sim a^2/G_N$. 
The remnant is therefore macroscopic, with a large number of microstates set by the original black-hole entropy.

Strictly speaking, whether the remnant is exactly stable, or evaporates on a parametrically longer timescale (e.g., the Poincaré time) through some residual process, is left open, and it is of cosmological importance \cite{PBH}.
What we have shown is that the standard Hawking radiation (i.e., the reinterpretation of the near-horizon vacuum as thermal radiation at large distances) is turned off.
The black hole may continue to lose its mass through other mechanisms, but the original paradox about how reinterpretation of vacuum as radiation can carry information is no longer a paradox.

The time-dependent shutoff observed in this work is invisible in standard analyses of Hawking radiation that assume a time-translation Killing vector.
A derivation tied to a dynamical collapse is necessary to expose the shutoff. 
This is one of the reasons why the scrambling-time cutoff has not been seen in the bulk of the Hawking-radiation literature, even though the trans-Planckian problem has been understood for decades.


\section*{Acknowledgements}

We thank Jil Le Bois, Johanna Borissova, Chong-Sun Chu, Norihiro Iizuka, Satoshi Iso, Hyun Jeong, and Hikaru Kawai for valuable discussions. 
P.M.H. is supported in part by the Ministry of Science and Technology, 
R.O.C. (NSTC 112-2112-M-002-024-MY3, NSTC 113-2112-M-002-040-MY2), 
and by National Taiwan University. 
W.H.S. is supported by the Special Postdoctoral Researcher (SPDR) Program at RIKEN.

\appendix


\end{document}